\documentclass[11pt,a4paper]{article}
\usepackage{amsmath}
\usepackage[mathcal]{eucal}
\usepackage{latexsym}
\usepackage{amssymb}
\begin{titlepage}
\title{Dirac consistency of the algebra of Hamiltonian constraints in reduced 4-D general relativity}
\author{Eyo Eyo Ita III}
\medskip
\input amssym.def
\input amssym.tex

\def \in{\indent}

\begin{document}
\maketitle
\bigskip
\centerline{Department of Applied Mathematics and Theoretical Physics} 
\smallskip
\centerline{Centre for Mathematical Sciences, University of Cambridge, Wilberforce Road}
\smallskip
\centerline{Cambridge CB3 0WA, United Kingdom}
\smallskip
\centerline{eei20@cam.ac.uk} 

\bigskip

\begin{abstract}
In this paper we provide an action related to a certain sector of general relativity where the algebra of Hamiltonian constraints forms a first class system.  This action is a Dirac-consistent stand-alone action with two physical degrees of freedom per point.  In this paper we provide the steps necessary to transform this new action to and from the associated sectors of the Ashtekar theory and a certain antecedent of the pure spin connection formulation by Capovilla, Dell and Jacobson.
\end{abstract}
\end{titlepage}
   
\section{Introduction}

The invariance of Einstein's theory of general relativity (GR) under general coordinate transformations is explicit at the covariant level of the theory, where space and time appear on equal footing.  In a canonical treatment one formulates the theory using variables defined on 3-dimensional spatial hypersurfaces which evolve in time.  This almost inevitably introduces a 3+1 splitting of the theory, and one must verify in the end that the original invariance has been preserved under this splitting.  In this paper we will probe this principle using a theory related to GR, specifically within the realm of time reparametrizations.  Let us consider a general transformation of coordinates $x\in{M}$, where $M$ is a 4-dimensional spacetime manifold
\begin{eqnarray}
\label{GENCOORD}
x^{\mu}\rightarrow{x^{\prime}}^{\mu}={x}^{\mu}+\xi^{\mu}(x).
\end{eqnarray}
\noindent
As shown in \cite{BLAGOJEVIC}, the form variation of a field $F(x)$, $\delta_0F(x)=F^{\prime}(x)-F(x)$ should be clearly distinguished from its total variation, $\delta{F}(x)\equiv{F}^{\prime}(x^{\prime})-F(x)$.  
Form variation and differentiation are commuting operations, and when $x^{\prime}-x$ is infinitesimally small, we get
\begin{eqnarray}
\label{GENCOORD1}
\delta{F}(x)\sim\delta_0F(x)+\xi^{\mu}\partial_{\mu}F(x).
\end{eqnarray}
\noindent
A scalar field $\varphi(x)\in{M}$ is a field which is invariant with respect to transformations (\ref{GENCOORD1}): $\varphi^{\prime}(x^{\prime})=\varphi(x)$.  As a consequence, the form variation of $\varphi$ is given by transformation law \cite{BLAGOJEVIC}
\begin{eqnarray}
\label{GENCOORD2}
\delta_0\varphi(x)=-\xi^{\mu}\partial_{\mu}\varphi(x).
\end{eqnarray}
Equation (\ref{GENCOORD}) as an infinitesimal general coordinate transformation defines the following vector field $\xi=\xi^{\mu}\partial_{\mu}\in{M}$ as realized in (\ref{GENCOORD1}).  The commutator of any two 
vector fields $\xi,\zeta\in{M}$ is given by the Lie bracket
\begin{eqnarray}
\label{LIE}
\bigl[\xi^{\mu}\partial_{\mu},\zeta^{\nu}\partial_{\nu}\bigr]=\bigl(\xi^{\mu}\partial_{\mu}\zeta^{\nu}-\zeta^{\mu}\partial_{\mu}\xi^{\nu}\bigr)\partial_{\nu},
\end{eqnarray}
\noindent
which defines a Lie algebra of general coordinate transformations.\par
\indent
To approach the question of whether there exists a formulation of GR where the Lie algebra (\ref{LIE}) can be realized at the canonical level, let us perform a 3+1 splitting of (\ref{LIE}) into purely spatial and temporal 
vector fields $\xi^{\mu}=(0,N^1,N^2,N^3)$ and $\xi^{\mu}=(N,0,0,0)$ for comparison.  This yields the following algebra
\begin{eqnarray}
\label{LIE3}
\bigl[N^i\partial_i,N^j\partial_j\bigr]=\bigl(M^i\partial_iN^j-N^i\partial_iM^j\bigr)\partial_j;\nonumber\\
\bigl[N^i\partial_i,N\partial_0\bigr]=(N^i\partial_iN)\partial_0-(N\dot{N}^i)\partial_i\nonumber\\
\bigl[M\partial_0,N\partial_0\bigr]=(M\dot{N}-N\dot{M})\partial_0.
\end{eqnarray}
\noindent
The Poisson algebra of hypersurface deformations for general relativity has been computed by Teitelboim \cite{TEITEL}
\begin{eqnarray}
\label{ALGEBRA121}
\{\vec{H}[\vec{N}],\vec{H}[\vec{M}]\}=H_k\bigl[N^{i}\partial^{k}M_i-M^{i}\partial^{k}N_i\bigr];\nonumber\\
\{H({N}),\vec{H}[\vec{N}]\}=H[N^{i}\partial_{i}{N}\bigr]\nonumber\\
\{H({N}),H({M})\}
=H_{i}[\bigl({N}\partial_{j}{M}-{M}\partial_{j}{N}\bigr)q^{ij}],
\end{eqnarray}
\noindent
where $H_{\mu}=(H,H_i)$ are the Hamiltonian and diffeomorphism constraints, and $q^{ij}$ are phase space dependent structure functions.  If one could make the identifications
\begin{eqnarray}
\label{IDENTEE}
H_{\mu}\sim\partial_{\mu}\longrightarrow{H}\sim\partial_0;~~H_i\sim\partial_i,
\end{eqnarray}
\noindent
then there would be an isomorphism between (\ref{LIE3}) and (\ref{ALGEBRA121}) with respect to purely spatial diffeomorphisms, which form a subalgebra of (\ref{LIE}).  However, equation (\ref{LIE3}) states that temporal diffeomorphisms should also form a subalgebra of (\ref{LIE}), which clearly is not the case in (\ref{ALGEBRA121}).\par
\indent
A direct implication of (\ref{ALGEBRA121}) is the nonexistence of a canonical formulation of GR, in the full theory, which evolves purely under the dynamics of the Hamiltonian constraint.\footnote{Another way to state this is that the Hamiltonian constraint $H$ does not form a first class system.  This is because the Poisson bracket of two Hamiltonian constraints yields a diffeomorphism constraint.  So (\ref{ALGEBRA121}) suggests that to be consistent, the diffeomorphism constraint $H_i$ must be part of the theory in addition to the Hamiltonian constraint $H$.}  In this paper we will propose an action $I_{Kin}$, which is directly related to a certain restricted sector of GR in a sense which we will make precise.  We will show that the temporal part of the algebra (\ref{LIE3}) is realized via Poisson brackets on $I_{Kin}$, which is the main result of this paper.  The question of whether $I_{Kin}$ is equivalent or not to GR is one which we will not address in this paper.  Rather, we will show that $I_{Kin}$ is a theory with 2 degrees of freedom per point on its reduced phase space, which is directly transformable into certain subspaces of the original theory of GR in ways which we will clearly demonstrate.\par
\indent  
The title of this paper refers to a `reduced' 4-dimensional GR theory, which is presented as $I_{Kin}$.  We will like to clarify that we have not shown in this paper that $I_{Kin}$ follows from full GR in the sense of a reduced phase space procedure, which involves solving constraints and gauge-fixing.  Rather, we will present $I_{Kin}$ as a restriction by hand from full GR to a certain subspace upon which our analysis will be carried out.  It will become clear that the action $I_{Kin}$ is still a stand-alone action irrespective of the issue of its precise relation GR.\footnote{So while we will not claim here that $I_{Kin}$ is the actual reduced phase space for gravity, we will present it as motivation for the prospect that such a formulation for GR, where this or something similar might perhaps be realizable, cannot be ruled out.}  The organization of this paper is as follows.  Sections 2 and 3 present the action $I_{Kin}$ as the starting point, which is a totally constrained system with a single constraint which we have named a Hamiltonian constraint.  We carry out the Dirac procedure for constrained systems, showing 
that $I_{Kin}$ is Dirac consistent at the classical level, and with a physical phase space having two degrees of freedom per point.  Sections 4 and 5 present the transformations which take $I_{Kin}$ to and from certain sectors of GR, specifically the restriction of full general relativity to the diagonal subspace of the Ashtekar and other variables and with no Gauss' law and diffeomorphism constraint.  Our main result will be to show that this is still consistent, even if it turns out to be the case that $I_{Kin}$ is not equivalent to GR.  Section 6 is a short discussion and conclusion of our results.  In Appendix A, we derive the set of configurations exhibiting the same 
features as $I_{Kin}$ as we have presented in this paper.\par
\indent
On a final note regarding index conventions for this paper, lowercase symbols $a,b,c,\dots$ from the beginning part of the Latin alphabet signify internal SO(3,C) indices, while those from the middle $i,j,k,\dots$ are spatial indices.  Both sets of indices will take values $1$, $2$ and $3$.  Greeek indices $\mu$, $\nu$ will denote spacetime indices, which take values $0$, $1$, $2$ and $3$.

\section{The starting action}

Consider the phase space $\Omega_{Kin}=(\Gamma_{Kin},P_{Kin})$ of a system with configuration and momentum space variables $\Gamma_{Kin}=(X,Y,T)$ and $P_{Kin}=(\Pi_1,\Pi_2,\Pi)$ defined on a 4-dimensional spacetime manifold of topology $M=\Sigma\times{R}$, where $\Sigma$ is a 3-dimensional spatial hypersurface with $R$ as the time direction.  The variables are in general complex, and the configuration space variables take on the 
ranges $-\infty<\vert{X}\vert,\vert{Y}\vert,\vert{T}\vert<\infty$.  The following mass dimensions have been assigned to the variables  
\begin{eqnarray}
\label{JACOBI8}
[\Pi_1]=[\Pi_2]=[\Pi]=1;~~[X]=[Y]=[T]=0.
\end{eqnarray}
\noindent
From these variables can be constructed the following kinematic phase space action for a totally constrained system
\begin{eqnarray}
\label{STARTINGACTIONN}
I=-{i \over G}\int{dt}\int_{\Sigma}d^3x\bigl(\Pi_1\dot{X}+\Pi_2\dot{Y}+\Pi\dot{T}\bigr)-iH[N],
\end{eqnarray}
\noindent
where $G$ is Newton's gravitational constant.  The function $H$ is smeared by an auxilliary field $N$, forming a Hamiltonian density $H[N]$ given by
\begin{eqnarray}
\label{STUFF}
H[N]=\int_{\Sigma}d^3xNUe^{-T/2}\Phi
\end{eqnarray}
\noindent
where the quantities in (\ref{STUFF}) are defined as follows.  First we have $\Phi$, given by
\begin{eqnarray}
\label{ALBOOT}
\Phi=\sqrt{\Pi(\Pi+\Pi_1)(\Pi+\Pi_2)}
\Bigl[\Bigl(k+e^T\Bigl({1 \over \Pi}+{1 \over {\Pi+\Pi_1}}+{1 \over {\Pi+\Pi_2}}\Bigr)\Bigr]
\end{eqnarray}
\noindent
where $k={\Lambda \over {a_0^3}}$ is a numerical constant.\footnote{We will identify $\Lambda$ as the cosmological constant, and $a_0$ is a numerical constant of mass dimension $[a_0]=1$.}  There are no spatial derivatives in any of the quantities in (\ref{ALBOOT}), and all spatial derivatives in the theory (\ref{STARTINGACTIONN}) are confined to the 
quantity $U$ in (\ref{STUFF}), given by
\begin{eqnarray}
\label{YOU}
U=\Bigl[1+e^{-T}\bigl((\partial_2Z)(\partial_3X)(\partial_1Y)-(\partial_3Y)(\partial_1Z)(\partial_2X)\bigr)\nonumber\\
+e^{-2X}(\partial_1Y)(\partial_1Z)+e^{-2Y}(\partial_2Z)(\partial_2X)+e^{-2Z}(\partial_3X)(\partial_3Y)\Bigr]^{1/2}
\end{eqnarray}
\noindent
with $Z=T-X-Y$.  We have defined
\begin{eqnarray}
\label{DEEFE}
\partial_1={\partial \over {\partial{y}^1}};~~\partial_2={\partial \over {\partial{y}^2}};~~\partial_3={\partial \over {\partial{y}^3}},
\end{eqnarray}
\noindent
where $y^1$, $y^1$ and $y^3$ are dimensionless spatial coordinates in 3-space $\Sigma$.\par
\indent
The canonical structure of (\ref{STARTINGACTIONN}) yields the following Poisson brackets amongst fundamental phase space variables
\begin{eqnarray}
\label{THEFOLLOWING}
\{{X}(x,t),{\Pi}_1(y,t)\}=\{{Y}(x,t),{\Pi}_2(y,t)\}=\{{T}(x,t),{\Pi}(y,t)\}=-iG\delta^{(3)}(x,y),
\end{eqnarray}
\noindent
with all other brackets vanishing.  Note that this induces the following canonical Poisson bracket between any two phase space function $f,g\in{C}^{\infty}(\Omega_{Kin})$
\begin{eqnarray}
\label{STARTINGACTIONN1}
\{f,g\}=\int_{\Sigma}d^3x\Bigl[{{\delta{f}} \over {\delta\Pi_1}}{{\delta{g}} \over {\delta{X}}}-{{\delta{g}} \over {\delta\Pi_1}}{{\delta{f}} \over {\delta{X}}}
+{{\delta{f}} \over {\delta\Pi_2}}{{\delta{g}} \over {\delta{Y}}}-{{\delta{g}} \over {\delta\Pi_2}}{{\delta{f}} \over {\delta{Y}}}
+{{\delta{f}} \over {\delta\Pi}}{{\delta{g}} \over {\delta{T}}}-{{\delta{g}} \over {\delta\Pi}}{{\delta{f}} \over {\delta{T}}}\Bigr].
\end{eqnarray}
\noindent
Since the Hamiltonian of (\ref{STARTINGACTIONN}) consists purely of a constraint proportional to $\Phi$, then it is appropriate to proceed with the Dirac analysis for totally constrained systems \cite{DIR}.\par
\indent
The velocity $\dot{N}$ does not appear in the starting action (\ref{STARTINGACTIONN}), which implies as a primary constraint the vanishing of its conjugate momentum $\Pi_N$
\begin{eqnarray}
\label{PRIMARY}
\Pi_N={{\delta{I}} \over {\delta\dot{N}}}=0.
\end{eqnarray}
\noindent
As a consistency condition we must require that $\Pi_N$ be preserved in time, which leads to the secondary constraint
\begin{eqnarray}
\label{PRIMARY1}
\dot{\Pi}_N={{\delta{I}} \over {\delta{N}}}=H=Ue^{-T/2}\Phi=0.
\end{eqnarray}
\noindent
We must then check for the preservation of (\ref{PRIMARY1}) in time, which is the same as checking for closure of the algebra of Hamiltonian (\ref{STUFF}) under Poisson brackets (\ref{STARTINGACTIONN1}).\par  
\indent

\section{Poisson algebra of the Hamiltonian constraint}

We will now compute the Poisson algebra of two Hamiltonians.  There exist phase space functions $q^I=q^I(\Omega_{Kin})$ such that the functional derivatives of (\ref{STUFF}) with respect to momentum space variables are weakly of the form
\begin{eqnarray}
\label{ALGEBRA1}
{{\delta{H}[N]} \over {\delta\Pi_1}}\sim{N}q^1;~~{{\delta{H}[N]} \over {\delta\Pi_2}}\sim{N}q^2;~~{{\delta{H}[N]} \over {\delta\Pi}}\sim{N}q^3
\end{eqnarray}
\noindent
where we have defined
\begin{eqnarray}
\label{GRADIENT4}
q^1=-Ue^{T/2}\sqrt{\Pi(\Pi+\Pi_1)(\Pi+\Pi_2)}\Bigl({1 \over {\Pi+\Pi_1}}\Bigr)^2\nonumber\\
q^2=-Ue^{T/2}\sqrt{\Pi(\Pi+\Pi_1)(\Pi+\Pi_2)}\Bigl({1 \over {\Pi+\Pi_2}}\Bigr)^2;\nonumber\\
q^3=-Ue^{T/2}\sqrt{\Pi(\Pi+\Pi_1)(\Pi+\Pi_2)}\Bigl[\Bigl({1 \over {\Pi}}\Bigr)^2
+\Bigl({1 \over {\Pi+\Pi_1}}\Bigr)^2+\Bigl({1 \over {\Pi+\Pi_2}}\Bigr)^2\Bigr].
\end{eqnarray}
For the configuration space the relevant contributions will arise from integration of $U$ by parts, which transfers the spatial gradients away from the variables whose functional derivatives are being evaluated.  For functional derivatives with respect to the `coordinate' $X$ we have that
\begin{eqnarray}
\label{RELEVANT}
-{{\delta{H}[M]} \over {\delta{X}}}=\partial_i(\eta^i_1M\Phi)+{1 \over U}Me^{T/2}\Phi\Bigl(-e^{-2X}(\partial_1Y)(\partial_1Z)+e^{-2Z}(\partial_3X)(\partial_3Y)\Bigr),
\end{eqnarray}
\noindent
where the following quantities have been defined
\begin{eqnarray}
\label{RELEVANT1}
\eta^1_1={1 \over {2U}}e^{T/2}\Bigl(-e^{-2X}(\partial_1Y)+e^{-T}(\partial_2X)(\partial_3Y)\Bigr);\nonumber\\
\eta^2_1={1 \over {2U}}e^{T/2}\Bigl(e^{-2Y}\partial_2(Z-X)-e^{-T}\bigl((\partial_3X)(\partial_1Y)+(\partial_3Y)(\partial_1Z)\bigr)\Bigr)\nonumber\\
\eta^3_1={1 \over {2U}}e^{T/2}\Bigl(e^{-2Z}(\partial_3Y)+e^{-T}(\partial_1Y)(\partial_2Z)\Bigr).
\end{eqnarray}
\noindent
For functional derivatives with respect to the `coordinate' $Y$ we have
\begin{eqnarray}
\label{RELEVANT2}
-{{\delta{H}[M]} \over {\delta{Y}}}=\partial_i(\eta^i_1M\Phi)+{1 \over U}Me^{T/2}\Phi\Bigl(-e^{-2Y}(\partial_2Z)(\partial_2X)+e^{-2Z}(\partial_3X)(\partial_3Y)\Bigr),
\end{eqnarray}
\noindent
where the following quantities have been defined
\begin{eqnarray}
\label{RELEVANT3}
\eta^1_2={1 \over {2U}}e^{T/2}\Bigl(e^{-2X}\partial_1(Z-Y)+e^{-T}\bigl((\partial_2Z)(\partial_3X)+(\partial_2X)(\partial_3Y)\bigr)\Bigr);\nonumber\\
\eta^2_2={1 \over {2U}}e^{T/2}\Bigl(-e^{-2Y}\bigl((\partial_2X)+(\partial_3X)(\partial_1Y)\bigr)\Bigr);\nonumber\\
\eta^3_2={1 \over {2U}}e^{T/2}\Bigl(e^{-2Z}(\partial_3X)-e^{-T}(\partial_1Z)(\partial_2X)\Bigr).
\end{eqnarray}
\noindent
For functional derivatives with respect to the `coordinate' $T$ we have
\begin{eqnarray}
\label{RELEVANT4}
-{{\delta{H}[M]} \over {\delta{T}}}=\partial_i(\eta^i_3M\Phi)-{1 \over {2U}}Me^{T/2}\Bigl(e^{-T}\bigl((\partial_2Z)(\partial_3X)(\partial_1Y)\nonumber\\
-(\partial_3Y)(\partial_1Z)(\partial_2X)\bigr)-e^{-2Z}(\partial_3X)(\partial_3Y),
\end{eqnarray}
\noindent
where the following quantities have been defined
\begin{eqnarray}
\label{RELEVANT5}
\eta^1_3={1 \over {2U}}\Bigl(e^{-2X}(\partial_1Y)-e^{-T}(\partial_2X)(\partial_3Y)\Bigr);\nonumber\\
\eta^2_3={1 \over {2U}}\Bigl(e^{-2Y}(\partial_2X)+e^{-T}(\partial_3X)(\partial_!Y)\Bigr);\nonumber\\
\eta^3_3=0.
\end{eqnarray}
\par
\indent
Let us now compute the individual terms contributing to the Poisson brackets between two Hamiltonian constraints smeared by auxilliary fields $N$ and $M$.  Using (\ref{ALGEBRA1}) and (\ref{RELEVANT}), (\ref{RELEVANT2}) 
and (\ref{RELEVANT4}) for the contribution due to $(\Pi_1,X)$ we have
\begin{eqnarray}
\label{ALGEBRA3}
\int_{\Sigma}d^3x\Bigl({{\delta{H}[N]} \over {\delta\Pi_1(x)}}{{\delta{H}[M]} \over {\delta{X}(x)}}
-{{\delta{H}[M]} \over {\delta\Pi_1(x)}}{{\delta{H}[N]} \over {\delta{X}(x)}}\Bigr)\nonumber\\
=\int_{\Sigma}d^3x\Bigl((Nq^1)\partial_i(\eta^i_1M\Phi)-(Mq^1)\partial_i(\eta^i_1N\Phi)\Bigr)
=\int_{\Sigma}d^3xq^1\eta^i_1\bigl(N\partial_iM-M\partial_iN\bigr)\Phi.
\end{eqnarray}
\noindent
Due to antisymmetry with respect to the difference of scalar functions, the only nontrivial contributions to (\ref{ALGEBRA3}) are from spatial derivatives acting on the functions $M$ and $N$.  Similarly for the $(\Pi_2,Y)$ contribution we have
\begin{eqnarray}
\label{ALGEBRA5}
\int_{\Sigma}d^3x\Bigl({{\delta{H}[N]} \over {\delta\Pi_2(x)}}{{\delta{H}[M]} \over {\delta{Y}(x)}}
-{{\delta{H}[M]} \over {\delta\Pi_2(x)}}{{\delta{H}[N]} \over {\delta{Y}(x)}}\Bigr)
=\int_{\Sigma}d^3xq^2\eta^i_2\bigl(N\partial_iM-M\partial_iN\bigr)\Phi.
\end{eqnarray}
\noindent
For the contribution to Poisson brackets due to $(\Pi,T)$ we have
\begin{eqnarray}
\label{ALGEBRA6}
\int_{\Sigma}d^3x\Bigl({{\delta{H}[N]} \over {\delta\Pi(x)}}{{\delta{H}[M]} \over {\delta{T}(x)}}
-{{\delta{H}[M]} \over {\delta\Pi(x)}}{{\delta{H}[N]} \over {\delta{T}(x)}}\Bigr)\nonumber\\
=\int_{\Sigma}d^3x\Bigl((Nq^3)\bigl(\partial_i(\eta^i_3M\Phi)+MC\bigr)-(Mq^1)\bigl(\partial_i(\eta^i_3N\Phi)+NC\bigr)\Bigr)\nonumber\\
=\int_{\Sigma}d^3x\Bigl((Nq^3)\partial_i(\eta^i_3M\Phi)-(Mq^3)\partial_i(\eta^i_3N\Phi)\Bigr)+\int_{\Sigma}d^3x\Bigl[(Nq^3)MC-(Mq^3)NC)\Bigr]
.
\end{eqnarray}
\noindent
The second integral on the last line on the right hand side of (\ref{ALGEBRA6}) vanishes, and the first integral simplifies to
\begin{eqnarray}
\label{ALGEBRA7}
\int_{\Sigma}d^3xq^3\eta^i_3\bigl(N\partial_iM-M\partial_iN\bigr)\Phi.
\end{eqnarray}
\noindent
Combining the results of (\ref{ALGEBRA7}), (\ref{ALGEBRA5}) and (\ref{ALGEBRA3}), we have that
\begin{eqnarray}
\label{ALGEBRA8}
\{H[N],H[M]\}=\int_{\Sigma}d^3xq^I\eta^j_I(N\partial_iM-M\partial_iN)\Phi=H[N,M],
\end{eqnarray}
\noindent
namely that the Poisson bracket of two Hamiltonian constraints is a Hamiltonian constraint with phase space dependent structure functions.  The result is that the classical Hamiltonian constraints algebra for (\ref{STARTINGACTIONN}) closes with no further constraints on the system.\par
\indent  
The classical constraints algebra of (\ref{STARTINGACTIONN}) closes, which implies that $\Omega_{Kin}$ constitutes a first class system.  A degree-of-freedom counting yields
\begin{eqnarray}
\label{DEEGREE}
3~(momentum)+3~(config.)-1~(First~Class~Constraint)\nonumber\\
-1~(Gauge-fixing)=4~phase~space~D.O.F.,
\end{eqnarray}
\noindent
which corresponds to two physical degrees of freedom per point.  The first class constraint is the Hamiltonian constraint $H$, and gauge-fixing of $I_{Kin}$ to its physical degrees of freedom involves factoring out the gauge orbits generated by $H$ in conjunction with making a choice of the auxilliary field $N$.  With two propagating degrees of freedom on its physical phase space, then we know that (\ref{STARTINGACTIONN}) is not a topological field theory.

\section{Relation of $I_{Kin}$ to general relativity}

There are at least two ways in which the starting action (\ref{STARTINGACTIONN}) is related to general relativity, which we will explain in the remainder of this paper.  (i) The first is the relation of $I_{Kin}$ to gravity in the Ashtekar variables (See e.g. \cite{ASH1}, \cite{ASH2} and \cite{ASH3}).  The Ashtekar action is given by  
\begin{eqnarray}
\label{ACTIONASH}
I_{Ash}=\int{dt}\int_{\Sigma}d^3x\Bigl[\widetilde{\sigma}^i_a\dot{A}^a_i+A^a_0D_i\widetilde{\sigma}^i_a\nonumber\\
-\epsilon_{ijk}N^i\widetilde{\sigma}^j_aB^k_a-{i \over 2}\underline{N}\epsilon_{ijk}\epsilon^{abc}\widetilde{\sigma}^i_a\widetilde{\sigma}^j_b\Bigl({\Lambda \over 3}\widetilde{\sigma}^k_c+B^k_c\Bigr)\Bigr],
\end{eqnarray}
\noindent
where $\widetilde{\sigma}^i_a$ is the densitized triad with $\underline{N}=N(\hbox{det}\widetilde{\sigma})^{-1/2}$ the densitized lapse function.  The configuration space variable $A^a_i$ is a gauge connection valued in SO(3,C).The fields $N^i$ and $A^a_0$ in (\ref{ACTIONASH}) are auxilliary fields smearing the Gauss' law and the diffeomorphism constraints.  Note that the constraints algebra of two Hamiltonian constraints 
from (\ref{ACTIONASH}) is given by \cite{ASH1}
\begin{eqnarray}
\label{KINEMAAT5}
\{H[M],H[N]\}=H_i[q^{ij}(M\partial_iN-N\partial_iM)],
\end{eqnarray}
\noindent
which has the same form as (\ref{ALGEBRA121}).  We will come back to this point later in this paper.\par
\indent  
(ii) The second way is the relation of $I_{Kin}$ to a certain action appearing in \cite{SPINCON}, which forms an intermediate step in obtaining the pure spin connection formulation $I_{CDJ}$ from Plebanski's theory of gravity \cite{PLEBANSKI}.  This action is 
\begin{eqnarray}
\label{NOOTED}
I_{(2)}=-{i \over G}\int{dt}\int_{\Sigma}d^3x\Bigl[{1 \over 8}\Psi_{ae}F^a_{\mu\nu}F^e_{\rho\sigma}\epsilon^{\mu\nu\rho\sigma}-i\eta\bigl(\Lambda+\hbox{tr}\Psi^{-1}\bigr)\Bigr],
\end{eqnarray}
\noindent
where $F^a_{\mu\nu}=\partial_{\mu}A^a_{\nu}-\partial_{\nu}A^a_{\mu}+f^{abc}A^b_{\mu}A^c_{\nu}$ is the curvature of a 4-dimensional $SO(3,C)$ connection $A^a_{\mu}$, and $\eta$ is a scalar density.  We would like to clarify 
that (\ref{NOOTED}) is not the final action proposed by Capovilla, Dell and Jacobson in \cite{SPINCON}.  The proposed action $I_{CDJ}$, which we will not display here, was obtained by elimination of 
the field $\Psi_{ae}$ from (\ref{NOOTED}), which we will refer to in this paper as the `CDJ action antecedent'.  We will now show that $I_{Kin}$ can be seen as a restriction of (\ref{ACTIONASH}) in conjunction with (\ref{NOOTED}) to certain sectors of phase space.\par
\indent
\subsection{Relation of $I_{Kin}$ to the CDJ action antecedent}
Consider the following transformations
\begin{eqnarray}
\label{TRANSFORMATION}
\Pi=a_0^3e^T\lambda_3;~~\Pi+\Pi_1=a_0^3e^T\lambda_1;~~\Pi+\Pi_2=a_0^3e^T\lambda_2
\end{eqnarray}
\noindent
for the momentum space variables $P_{Kin}$, and
\begin{eqnarray}
\label{TRANSFORMATION1}
X=\hbox{ln}\Bigl({{a_1} \over {a_0}}\Bigr);~~Y=\hbox{ln}\Bigl({{a_2} \over {a_0}}\Bigr);~~T=\hbox{ln}\Bigl({{a_1a_2a_3} \over {a_0^2}}\Bigr)
\end{eqnarray}
\noindent
for the configuration space variables $\Gamma_{Kin}$, where $a_0$ is a numerical constant of mass dimension $[a_0]=1$.  Note that the new coordinates have the ranges $0<\vert{a_a}\vert<\infty$ for $a=1,2,3$, which forms a 3-dimensional functional manifold per point with the origin $a_a=0$ missing.  Let us also make the definitions
\begin{eqnarray}
\label{TRANSFORMATION2}
x^1={{y^1} \over {a_0}};~~x^2={{y^2} \over {a_0}};~~x^3={{y^3} \over {a_0}}
\end{eqnarray}
\noindent
with $y^1$, $y^2$ and $y^3$ the dimensionless spatial coordinates in (\ref{DEEFE}), whence $[x^i]=-1$.  Substitution of (\ref{TRANSFORMATION1}) and (\ref{TRANSFORMATION2}) into (\ref{YOU}) yields
\begin{eqnarray}
\label{TRANSFORMATION3}
U=(a_1a_2a_3)^{-1}\biggl[(a_1a_2a_3)^2+(\partial_2a_3)(\partial_3a_1)(\partial_1a_2)\nonumber\\
-(\partial_3a_2)(\partial_1a_3)(\partial_2a_1)+a_2a_3(\partial_1a_2)(\partial_1a_3)+a_3a_1(\partial_2a_3)(\partial_2a_1)\nonumber\\
+a_1a_2(\partial_3a_1)(\partial_3a_2)\biggr]^{1/2}=(\hbox{det}A)^{-1}(\hbox{det}B)^{1/2},
\end{eqnarray}
\noindent
from which one recognizes $U$ as the square root of the determinant of the magnetic field $B^i_a$ for a diagonal connection $A^a_i=diag(a_1,a_2,a_3)$, with the leading order term in $(\hbox{det}A)$ factored out.  In matrix form this is given by
\begin{displaymath}
a^a_i=
\left(\begin{array}{ccc}
a_1 & 0 & 0\\
0 & a_2 & 0\\
0 & 0 & a_3\\
\end{array}\right);~~
b^i_a=
\left(\begin{array}{ccc}
a_2a_3 & -\partial_3a_2 & \partial_2a_3\\
\partial_3a_1 & a_3a_1 & -\partial_1a_3\\
-\partial_2a_1 & \partial_1a_2 & a_1a_2\\
\end{array}\right)
.
\end{displaymath}
\noindent
Substitution of (\ref{TRANSFORMATION}), (\ref{TRANSFORMATION1}) and(\ref{TRANSFORMATION3}) into (\ref{STARTINGACTIONN}) yields
\begin{eqnarray}
\label{REPEAT}
I=-{i \over G}\int{dt}\int_{\Sigma}d^3x\Bigl(\lambda_1a_2a_3\dot{a}_1+\lambda_2a_3a_1\dot{a}_2+\lambda_3a_1a_2\dot{a}_3\nonumber\\
-iN(\hbox{det}b)^{1/2}\sqrt{\lambda_1\lambda_2\lambda_3}\Bigl(\Lambda+{1 \over {\lambda_1}}+{1 \over {\lambda_2}}+{1 \over {\lambda_3}}\Bigr).
\end{eqnarray}
\noindent
Equation (\ref{REPEAT}) is nothing other than the 3+1 decomposition of (\ref{NOOTED}) with the Gauss' law constraint missing, with a phase space restricted to diagonal 
variables $A^a_i=diag(a_1,a_2,a_3)$ and $\Psi_{ae}=diag(\lambda_1,\lambda_2,\lambda_3)$.  Equation (\ref{REPEAT}) can be seen as the result of choosing $A^a_0=0$ at the level of the action (\ref{NOOTED}), which in certain interpretations corresponds to a gauge-fixing choice.  In this sense the possibility exists that (\ref{REPEAT}), while shown under the guise of (\ref{STARTINGACTIONN}) to be a Dirac consistent theory, could conceivably be a different theory from (\ref{NOOTED}) in actuality.\par
\indent
The action (\ref{REPEAT}) has the peculiar feature that its canonical one form does not have any spatial derivatives.  But there are spatial derivatives contained in the factor $(\hbox{det}b)^{1/2}$ in its 
Hamiltonian, and therefore (\ref{REPEAT}) is not a minisuperspace theory.  The canonical one-form in (\ref{REPEAT}) can be seen as the restriction to diagonal variables of the object
\begin{eqnarray}
\label{TRANSFORMATION5}
\boldsymbol{\theta}=\int_{\Sigma}d^3x\Psi_{ae}B^i_e\dot{A}^a_i\biggl\vert_{diag(\Psi;A)}.
\end{eqnarray}
\noindent
It so happens, since all spatial derivatives from the magnetic field $B^i_a$ occur in the off-diagonal matrix positions when $\dot{A}^a_i$ is diagonal, that the contraction with a diagonal matrix $\Psi_{ae}=diag(\lambda_1,\lambda_2,\lambda_3)$ annihilates these derivative terms.  There are six distinct configurations of $A^a_i$ which exhibit this feature, and we will refer to these configurations as `quantizable configurations' of configuration 
space $\Gamma_q$.  The configurations $\Gamma_q$ are given by
\begin{displaymath}
a^a_i=
\left(\begin{array}{ccc}
a^1_1 & 0 & 0\\
0 & a^2_2 & 0\\
0 & 0 & a^3_3\\
\end{array}\right),~
\left(\begin{array}{ccc}
a^1_1 & 0 & 0\\
0 & 0 & a^2_3\\
0 & a^3_2 & 0\\
\end{array}\right),~
\left(\begin{array}{ccc}
0 & a^2_1 & 0\\
a^1_2 & 0 & 0\\
0 & 0 & a^3_3\\
\end{array}\right),
\end{displaymath}
\begin{displaymath}
\left(\begin{array}{ccc}
0 & a^2_1 & 0\\
0 & 0 & a^3_2\\
a^1_3 & 0 & 0\\
\end{array}\right),~
\left(\begin{array}{ccc}
0 & 0 & a^3_1\\
a^1_2 & 0 & 0\\
0 & a^2_3 & 0\\
\end{array}\right),~
\left(\begin{array}{ccc}
0 & 0 & a^3_1\\
0 & a^2_2 & 0\\
a^1_3 & 0 & 0\\
\end{array}\right)
\in\Gamma_q,
\end{displaymath}
\noindent
namely the set of connections $a^a_i$ having three nonvanishing elements, and with $\hbox{det}a\neq{0}$.  The proof of this is provided in Appendix A.  Note that the same Dirac procedure as in sections 2 and 3 can be applied to each of the six configurations $\Gamma_q$ just as for the diagonal one considered.  Hence there are six separate sectors of a theory of $I_{Kin}$ which can be studied.

\section{Relation of $I_{Kin}$ to the Ashtekar variables}

To see the relation of (\ref{STARTINGACTIONN}) to the Ashtekar variables, let us perform a canonical analysis at the level of (\ref{REPEAT}).  The momenta canonically conjugate to the (diagonal) connection are given 
by $p_a=\delta{I}_{Kin}/\delta\dot{a}_a$, namely
\begin{eqnarray}
\label{KINEMAAT2}
p_1=\lambda_1a_2a_3;~~p_2=\lambda_2a_3a_1;~~p_3=\lambda_3a_1a_2.
\end{eqnarray}
\noindent
Let us now substitute (\ref{KINEMAAT2}) into the Hamiltonian density of (\ref{REPEAT}).  This yields
\begin{eqnarray}
\label{KINEMAAT4}
H=(\hbox{det}b)^{1/2}\sqrt{\lambda_1\lambda_2\lambda_3}\Bigl(\Lambda+{1 \over {\lambda_1}}+{1 \over {\lambda_2}}+{1 \over {\lambda_3}}\Bigr)\nonumber\\
=(\hbox{det}b)^{1/2}{{\sqrt{p_1p_2p_3}} \over {(a_1a_2a_3)}}\Bigl(\Lambda+{{a_2a_3} \over {p_1}}+{{a_3a_1} \over {p_2}}+{{a_1a_2} \over {p_3}}\Bigr)\nonumber\\=
U(p_1p_2p_3)^{-1/2}\bigl(\Lambda{p}_1p_2p_3+p_1p_2(a_1a_2)+p_2p_3(a_2a_3)+p_3p_1(a_3a_1)\bigr),
\end{eqnarray}
\noindent
with $U$ given by (\ref{TRANSFORMATION3}).  Substitution of (\ref{KINEMAAT2}) back into (\ref{REPEAT}) yields the action
\begin{eqnarray}
\label{KINEMAAT3}
I[p,a]=\int{dt}\int_{\Sigma}d^3xp_a\dot{a}^a-iNU(\hbox{det}p)^{-1/2}H,
\end{eqnarray}
\noindent
with $U$ as defined as in (\ref{TRANSFORMATION3}) and with
\begin{eqnarray}
\label{KINEMAAT31}
H=\Lambda{p}_1p_2p_3+p_1p_2(a_1a_2)+p_2p_3(a_2a_3)+p_3p_1(a_3a_1).
\end{eqnarray}
\noindent
In the case where the connection $A^a_i$ is spatially homogeneous, all derivatives in $U$ vanish and (\ref{KINEMAAT3}) reduces to a diagonal Bianchi I model.  But $a_a=a_a(x)$ in general contains three degrees of freedom per point, corresponding to three free functions of position and time.  The spatial derivatives $\partial_ia$ in general are nonzero, and therefore (\ref{KINEMAAT3}), as well as (\ref{REPEAT}), are not minisuperspace theories.\par
\indent 
The action (\ref{ACTIONASH}) with the Gauss' law and diffeomorphism constraints removed by hand is given by\footnote{The removal of Gauss' law and the diffeomorphism constraints by hand can in certain interpretations be seen as a gauge-fixing choice $N^i=A^a_0=0$ at the level of the action (\ref{ACTIONASH}).  This implies in certain interpretations that (\ref{NOACTIONASH}) and (\ref{ACTIONASH}) most likely are two inequivalent theories.}
\begin{eqnarray}
\label{NOACTIONASH}
I=\int{dt}\int_{\Sigma}d^3x\Bigl[\widetilde{\sigma}^i_a\dot{A}^a_i-{i \over 2}\underline{N}\epsilon_{ijk}\epsilon^{abc}\widetilde{\sigma}^i_a\widetilde{\sigma}^j_b\Bigl({\Lambda \over 3}\widetilde{\sigma}^k_c+B^k_c\Bigr)\Bigr].
\end{eqnarray}
\noindent
Recall that the Poisson bracket between two Hamiltonian constraints is a diffeomorphism constraint as in (\ref{KINEMAAT5}).  Since there is no diffeomorphism constraint contained in (\ref{NOACTIONASH}), then this action in its present form cannot be Dirac consistent in the full theory.  But suppose that we restrict (\ref{NOACTIONASH}) to the subspace of spatially inhomogeneous diagonal variables
\begin{displaymath}
\widetilde{\sigma}^i_a=
\left(\begin{array}{ccc}
p_1(x) & 0 & 0\\
0 & p_2(x) & 0\\
0 & 0 & p_3(x)\\
\end{array}\right);~~
A^a_i=
\left(\begin{array}{ccc}
a_1(x) & 0 & 0\\
0 & a_2(x) & 0\\
0 & 0 & a_3(x)\\
\end{array}\right)
\end{displaymath}
\noindent
with 3 D.O.F. per point.  Then for $\widetilde{\sigma}^i_a=\delta^i_ap_a$ and $A^a_i=\delta^a_ia_a$ with no summation over $a$, the action (\ref{NOACTIONASH}) is given by
\begin{eqnarray}
\label{NOACTIONASH1}
I=\int{dt}\int_{\Sigma}d^3x\Bigl[\widetilde{\sigma}^i_a\dot{A}^a_i-iN(\hbox{det}\widetilde{\sigma})\bigl(\Lambda+(\widetilde{\sigma}^{-1})^a_iB^i_a\Bigr]\biggl\vert_{Diag(A;\widetilde{\sigma})}\nonumber\\
=\int{dt}\int_{\Sigma}d^3x\Bigl[p_a\dot{a}_a-i\underline{N}\bigl(\Lambda{p}_1p_2p_3+p_1p_2(a_1a_2)+p_2p_3(a_2a_3)+p_3p_1(a_3a_1)\bigr)\Bigr].
\end{eqnarray}
\noindent
Equation (\ref{NOACTIONASH1}) can be seen as a special case of (\ref{KINEMAAT3}) when $U=1$, with $U$ as defined in (\ref{TRANSFORMATION3}).  Since all spatial derivatives in (\ref{STARTINGACTIONN}) and in (\ref{KINEMAAT3}) are 
confined $U$, then (\ref{NOACTIONASH}) on diagonal variables, even when not spatially homogeneous, is no more general than a minisuperspace theory.\footnote{This is because there are no spatial derivatives in (\ref{NOACTIONASH1}), which moreover is Dirac inconsistent unless the variables are chosen to be spatially homogeneous.  The spatial derivatives in (\ref{NOACTIONASH1}) have dropped out for the same reason that they drop out of the canonical one form of (\ref{REPEAT}).  However recall that (\ref{REPEAT}) still has spatial derivatives contained in $(\hbox{det}b)^{1/2}$ which multiplies the lapse function $N$, whereas (\ref{NOACTIONASH}) and (\ref{NOACTIONASH1}) do not.}  Therefore the restriction of the Ashtekar theory to diagonal variables yields a theory not having spatial derivatives, which is essentially the same as a minisuperspace theory.  So the action (\ref{STARTINGACTIONN}) is equivalent with the diagonally restricted Ashtekar theory only in minisuperspace, for the special case $U=1$.  In the full theory where $U\neq{0}$, then this is not so and 
while (\ref{NOACTIONASH}) is Dirac-inconsistent, equation (\ref{STARTINGACTIONN}) is a Dirac consistent theory as we have demonstrated.  So these two actions are definitely not equivalent on the subspace of diagonal variables in the general case.  This then brings in the question of whether there exists action for (\ref{STARTINGACTIONN}) which for $U\neq{1}$ constitutes analogue of the diagonally restricted version of (\ref{NOACTIONASH}), such that the action is not inconsistent in the full theory as is (\ref{NOACTIONASH}).  We will relegate the writing down of the desired action to the discussion section of this paper.\par
\indent
\subsection{Resolution of the disparity between minisuperspace and the full theory}

We will now revisit the question of whether there exists a consistent action analogous to (\ref{NOACTIONASH}), which can be interpreted as the antecedent of the Dirac-consistent action (\ref{STARTINGACTIONN}).  The arguments of the previous section show that in minisuperspace where $U=1$, equation (\ref{STARTINGACTIONN}) can be obtained by removing the Gauss' law and diffeomorphism constraints and restricting (\ref{ACTIONASH}) to diagonal 
variables.  Moreover, (\ref{ACTIONASH}) leads via these restrictions initially to (\ref{NOACTIONASH}), which is not Dirac consistent in the full theory.  Since (\ref{STARTINGACTIONN}) is a Dirac consistent theory in the full theory, then a pertinent question regards the mechanism by which the Dirac-inconsistent (\ref{NOACTIONASH}) can become associated with the a Dirac-consistent (\ref{STARTINGACTIONN}) in the general case $U\neq{1}$.\par
\indent
The root cause for the disparity apparently resides in the term $U$, which contains all spatial derivatives of the theory.  Recall that $U$ is contained in (\ref{STARTINGACTIONN}) but is not contained in (\ref{NOACTIONASH}).  There is a certain transformation known as the CDJ Ansatz\footnote{This can be seen as the spatial restriction of one of the equations of motion arising in Plebanski's theory of gravity \cite{PLEBANSKI}.}
\begin{eqnarray}
\label{REPEATEDIT1}
\widetilde{\sigma}^i_a=\Psi_{ae}B^i_e,
\end{eqnarray}
\noindent
where $\Psi_{ae}=\Psi_{(ae)}\in{SO}(3,C)\times{SO}(3,C)$ is symmetric, transforms (\ref{ACTIONASH}) into the action (\ref{NOOTED}) when $(\hbox{det}B)\neq{0}$ and $(\hbox{det}\Psi)=0$.  Let us examine the implication 
of (\ref{REPEATEDIT1}) for (\ref{KINEMAAT3}) and (\ref{REPEAT}), the `reduced' versions of (\ref{ACTIONASH}) and (\ref{NOOTED}) which follow from (\ref{STARTINGACTIONN}).  Note that (\ref{REPEAT}) can be written as
\begin{eqnarray}
\label{REPEATEDIT}
I=-{i \over G}\int{dt}\int_{\Sigma}d^3x\Bigl[\Psi_{ae}B^i_e\dot{A}^a_i\nonumber\\
-iN(\hbox{det}B)^{1/2}\sqrt{\hbox{det}\Psi}\bigl(\Lambda+\hbox{tr}\Psi^{-1}\bigr)\Bigr]\biggl\vert_{diag(A);diag(\Psi)},
\end{eqnarray}
\noindent
with phase space restrictions $\Psi_{ae}=\delta_{ae}\Psi_{aa}\equiv\delta_{ae}\lambda_e$ and $A^a_i=\delta^a_ia_a$ to diagonal variables.  The unrestricted versrion of (\ref{REPEATEDIT}), namely where the variables can be nondiagonal, is simply the 3+1 decomposition of (\ref{NOOTED}) with the Gauss' law constraint removed.  An easy way to see this is to look at the integrand of the canonical one form.  First use the following definitions for the components of the curvature
\begin{eqnarray}
\label{ONEFORMM}
B^i_a={1 \over 2}\epsilon^{ijk}F^a_{jk};~~F^a_{0i}=\dot{A}^a_i-D_iA^a_0,
\end{eqnarray}
\noindent 
where $D_iv_a=\partial_iv_a+f_{abc}A^b_iv_c$ is the $SO(3,C)$ covariant derivative of the $SO(3,C)$-valued vector $v_a$.  Then defining $\epsilon^{ijk}\equiv\epsilon^{0ijk}$ and using the symmetries of the 4-D 
epsilon symbol $\epsilon^{\mu\nu\rho\sigma}$, we have
\begin{eqnarray}
\label{ONEFORM1}
\Psi_{(ae)}B^i_e\dot{A}^a_i={1 \over 2}\Psi_{(ae)}\epsilon^{ijk}F^e_{jk}(F^a_{0i}+D_iA^a_0)\nonumber\\
={1 \over 8}\Psi_{ae}F^a_{\mu\nu}F^e_{\rho\sigma}\epsilon^{\mu\nu\rho\sigma}+\Psi_{(ae)}B^i_eD_iA^a_0.
\end{eqnarray}
\noindent
The first term on the right hand side of (\ref{ONEFORM1}) is the same as the first term of (\ref{NOOTED}), which includes the Gauss' constraint.  The second term of (\ref{ONEFORM1}) removes this Gauss' constraint, which can be obtained by integration by parts with discarding of boundary terms $\Psi_{(ae)}B^i_eD_iA^a_0\rightarrow-A^a_0B^i_eD_i\Psi_{(ae)}$.  The same holds true on the diagonally restricted subspace of this.\par
\indent
Equation (\ref{REPEATEDIT}) is the same as the Dirac consistent theory (\ref{STARTINGACTIONN}) after the redesignation of variables (\ref{TRANSFORMATION}) and (\ref{TRANSFORMATION1}).  But substitution of (\ref{REPEATEDIT1}) in conjunction with restriction to diagonal variables transforms (\ref{NOACTIONASH}) 
into (\ref{REPEATEDIT}).  Since (\ref{NOACTIONASH}) under (\ref{REPEATEDIT1}) transforms, upon restriction to diagonal variables, into (\ref{REPEAT}), and (\ref{REPEAT}) transforms via canonical transformation into (\ref{KINEMAAT3}), then it follows that (\ref{REPEATEDIT1}) is a noncanonical transformation.  The conclusion is that this noncanonical transformation, in conjunction with a restriction to diagonal variables (or any of the quantiable 
configurations $\Gamma_q$) is what is necessary to make a Dirac consistent theory out of the reduction (as we have defined it in this paper) of (\ref{ACTIONASH}).  A way to see this is that equation (\ref{REPEATEDIT1}) contains spatial derivatives on the right hand side in $B^i_a$, whereas there are no spatial derivatives explicitly present on the left hand side.  It is precisely these derivatives from $B^i_a$ which make the difference between a Dirac-consistent  full theory of (\ref{STARTINGACTIONN}) and a Dirac-inconsistent full-theory of (\ref{NOACTIONASH}).\footnote{The latter being Dirac-consistent only in minisuperspace.}

\section{Conclusion and discussion}

The main aim of this paper at presenting an action (\ref{STARTINGACTIONN}) which realizes the Lie subalgebra of temporal coordinate transformations (\ref{LIE3}) has been carried out.\footnote{This is notwithstanding the fact that there are phase space structure functions appearing in (\ref{ALGEBRA8}) which still need to be interpreted.}  We have presented an action $I_{Kin}$ in equation (\ref{STARTINGACTIONN}) which has been shown to be Dirac consistent at the classical level and to exhibit two physical degrees of freedom per point.  We have shown the relation of $I_{Kin}$ to two formulations of general relativity, namely the Ashtekar variables and a certain antecedent of the CDJ pure spin connection formulation in \cite{SPINCON}.  In basic terms, the action $I_{Kin}$ can be seen as a restriction of the actions of these formulations to diagonal variables where the Gauss' law and diffeomorphism constraints have been removed by hand.  While this is strictly speaking, not technically rigorous as a gauge-fixing procedure, the associated action $I_{Kin}$ is still nevertheless a stand-alone action in the full theory and consistent in the Dirac sense.\footnote{For an analogy, the action $I_{Ash}$ for GR in Ashtekar variables \cite{ASH1} can be obtained from Plebanski's $I_{Pleb}$ action \cite{PLEBANSKI} in the so-called time gauge, which sets three degrees of freedom coresponding to the choice of a Lorentz frame to zero.  But even though $I_{Ash}\subset{I}_{Pleb}$ is a restriction of Plebanski's action to this specialized sector, the Ashtekar action is still self-consistent in the Dirac sense and is a stand-alone action irrespective of the issue of its equivalence with $I_{Pleb}$.}  Hence we would like (\ref{STARTINGACTIONN}) to serve as a motivation for putting in place a rigorously correct gauge-fixing procedure for full GR.  The issue of equivalence of the theories in light of the restrictions, or gauge-choices in certain interpretations, is one which we have reserved for addressal in a subsequent paper.\par
\indent

\indent
\section{Appendix A: Quantizable configurations of configuration space}

\noindent
We have shown that the kinematic phase space action (\ref{STARTINGACTIONN}) can be seen as the diagonal subspace of an action appearing in \cite{SPINCON} except with the Gauss' constraint missing.  But we have shown that this action is Dirac consistent for a diagonal connection.  However, (\ref{NOACTIONASH}) is Dirac consistent only in minisuperspace for a diagonal connection.  This leads to the question of whether there are any additional configurations analogous to the diagonal case arising from (\ref{STARTINGACTIONN}), which are Dirac consistent.\par
\indent
The reason why (\ref{STARTINGACTIONN}) rather than (\ref{REPEAT}) is in suitable form for canonical analysis is because (\ref{REPEAT}) is not in canonical form.  This can be seen from the fact that the variation of its canonical one form, even for the case of a diagonal connection
\begin{eqnarray}
\label{THISCANBE}
\delta\Bigl(\int_{\Sigma}d^3x\lambda_ab^i_a\delta{a}^a_i\Bigr)\biggl\vert_{diag(A)}\nonumber\\
=\int_{\Sigma}d^3x\Bigl[(a_2a_3){\delta\lambda_1}\wedge{\delta{a}_1}+\lambda_1{\delta(a_2a_3)}\wedge{\delta{a}_1}\Bigr]+Cyclic~Perms,
\end{eqnarray}
\noindent
does not yield a closed symplectic 2-form owing to the second term on the right hand side of (\ref{THISCANBE}).  This difficulty is compounded in the more general case where one is not limited to diagonal variables, which brings spatial derivatives into the symplectic 2-form
\begin{eqnarray}
\label{CANON}
\delta\boldsymbol{\theta}_{general}=\delta\Bigl(\int_{\Sigma}d^3x\lambda_fb^i_f\delta{a}^f_i\Bigr)
=\int_{\Sigma}d^3x\Bigl[b^i_f{\delta\lambda_f}\wedge{\delta{a}^f_i}+\lambda_f({\epsilon^{ijk}D_j\delta{a}^f_k})\wedge{\delta{a}^f_i}\Bigr].
\end{eqnarray}
\noindent
Equation (\ref{CANON}) is not a symplectic two form $\boldsymbol{\Omega}_{general}$ of canonical form $\boldsymbol{\Omega}=\delta(p\delta{q})={\delta{p}}\wedge{\delta{q}}$, and is not suitable for quantization.  The configuration space part of $\boldsymbol{\theta}_{Kin}$ splits into two contributions $b^i_f\delta{a}^f_i=m_f+n_f$, where
\begin{eqnarray}
\label{CANON1}
m_f=\epsilon^{ijk}(\partial_ja^f_k)\delta{a}^f_i;~~n_f={1 \over 2}\epsilon^{ijk}f_{fgh}a^g_ja^h_k\delta{a}^f_i.
\end{eqnarray}
\noindent
Note that $m_f$ contains spatial gradients of $a^f_i$, while $n_f$ is free of spatial gradients.  We will see that a sufficient condition for (\ref{CANON}) to admit a canonical structure on $\Omega_{Kin}$ is that the second term on the right hand side of (\ref{CANON}) vanishes, which is tantamount to the requirement that $m_f$ in (\ref{CANON1}) be zero for all $f$.  Let us determine the configurations $a^f_i$ for which this is the case by expanding $m_f$ and rearranging the terms into the following form
\begin{eqnarray}
\label{CANON2}
m_f=(\partial_2a^f_3-\partial_3a^f_2)\delta{a}^f_1
+(\partial_3a^f_1-\partial_1a^f_3)\delta{a}^f_2
+(\partial_1a^f_2-\partial_2a^f_1)\delta{a}^f_3\nonumber\\
=\bigl((\delta{a}^f_2)\partial_3-(\delta{a}^f_3)\partial_2\bigr)a^f_1
+\bigl((\delta{a}^f_3)\partial_1-(\delta{a}^f_1)\partial_3\bigr)a^f_2
+\bigl((\delta{a}^f_1)\partial_2-(\delta{a}^f_2)\partial_1\bigr)a^f_3.
\end{eqnarray}
\noindent
From (\ref{CANON2}) it is clear that a sufficient condition for $m_f=0$ is that all except three matrix elements of $a^f_i$ be zero, with the nonzero elements such that no two appear in the same row or column.  In other words, we must have $(\hbox{det}a^f_i)\neq{0}$, which restricts the connection to one of the six forms
\begin{displaymath}
a^a_i=
\left(\begin{array}{ccc}
a^1_1 & 0 & 0\\
0 & a^2_2 & 0\\
0 & 0 & a^3_3\\
\end{array}\right),~
\left(\begin{array}{ccc}
a^1_1 & 0 & 0\\
0 & 0 & a^2_3\\
0 & a^3_2 & 0\\
\end{array}\right),~
\left(\begin{array}{ccc}
0 & a^2_1 & 0\\
a^1_2 & 0 & 0\\
0 & 0 & a^3_3\\
\end{array}\right),
\end{displaymath}
\begin{displaymath}
\left(\begin{array}{ccc}
0 & a^2_1 & 0\\
0 & 0 & a^3_2\\
a^1_3 & 0 & 0\\
\end{array}\right),~
\left(\begin{array}{ccc}
0 & 0 & a^3_1\\
a^1_2 & 0 & 0\\
0 & a^2_3 & 0\\
\end{array}\right),~
\left(\begin{array}{ccc}
0 & 0 & a^3_1\\
0 & a^2_2 & 0\\
a^1_3 & 0 & 0\\
\end{array}\right)
\in\Gamma_q,
\end{displaymath}
\noindent
where $\Gamma_q$ defines what we will refer to as the quantizable configurations of configuration space.  Hence for $a^f_i\in\Gamma_q$, we have that $m_f=0$, and that $n_f$ is given by
\begin{eqnarray}
\label{CANON3}
n_f={1 \over 2}\epsilon^{ijk}f_{fgh}a^g_ja^h_k\delta{a}^f_i=(\hbox{det}a)(a^{-1})^i_f\delta{a}^f_i.
\end{eqnarray}
\noindent
It is not difficult to see that each of the six configurations $\Gamma_q$ leads to a Dirac consistent theory as the diagonal sector we have illustrated in this paper.  This constitutes six distinct sectors of the full theory (and not minisuperspace) of reduced general relativity that can be studied.

\end{document}